\documentclass{sig-alternate-05-2015}
\usepackage{graphicx}
\usepackage{amssymb}
\usepackage[noend]{algorithmic}
\usepackage{algorithm}
\usepackage{textcomp}
\usepackage{listings}
\usepackage[usenames,dvipsnames]{xcolor}
\usepackage{subcaption}

\setcopyright{acmcopyright}

\numberofauthors{1}
\author{Arindam Nandi$^\beta$, Ying Yang$^\beta$, Oliver Kennedy$^\beta$\\[0.5mm]
Boris Glavic$^i$, Ronny Fehling$^\alpha$, Zhen Hua Liu$^o$, Dieter Gawlick$^o$\\[2mm]
\affaddr{University at Buffalo$^\beta$} 
\hspace{5mm} \affaddr{Illinois Institute of Technology$^i$}
\hspace{5mm} \affaddr{Airbus$^\alpha$}
\hspace{5mm} \affaddr{Oracle$^o$}\\[2mm]
\email{\{arindamn, yyang25, okennedy\}@buffalo.edu} \hspace{5mm} \email{bglavic@iit.edu}\\[0.5mm]
\email{ronny.fehling@airbus.com} \hspace{5mm} \email{\{zhen.liu,dieter.gawlick\}@oracle.com}
}

\title{Mimir: Bringing CTables into Practice\titlenote{The first two authors contributed equally and should be considered a joint first author}}

\newtheorem{example}{Example}

\newcommand{\comprehension}[2]{\left\{\;{#1}\;|\;{#2}\;\right\}}
\newcommand{\tuple}[1]{\left<\;{#1}\;\right>}
\newcommand{\ordefn}{\;|\;}

\newcommand{\projection}{\pi}

\newcommand{\ctifthenelse}[3]{\textbf{if}\ {#1}\ \textbf{then}\ {#2}\ \textbf{else}\ {#3}}

\newcommand{\isnull}[1]{{#1}\;\texttt{is\;null}}
\newcommand{\ROWID}{\texttt{ROWID}}
\newcommand{\evallazy}[1]{[[{#1}]]_{lazy}}

\newcommand{\nonDetColumn}[1]{\textcolor{Blue}{{#1}}}

\newcommand{\applyBind}[2]{#2[#1]}

\newcommand{\ccomment}[1]{{\small\texttt{/*} #1 \texttt{*/}}}

\newcommand{\D}{\mathcal D}

\newcommand{\F}{\mathcal F}

\newcommand{\tinysection}[1]{\noindent \textbf{#1.}~}

\toappear{}

\begin{document}
\maketitle

\begin{abstract}
The present state of the art in analytics requires high upfront investment of human 
effort and computational resources to curate datasets, even before the first 
query is posed.  So-called pay-as-you-go data curation techniques allow
these high costs  to be spread out, first by enabling queries over uncertain and 
incomplete data, and then by assessing the quality of the query results.  We describe
the design of a system, called Mimir, around a recently introduced class of 
probabilistic pay-as-you-go data cleaning operators called Lenses.  Mimir wraps
around any deterministic database engine using JDBC, extending it with
support for probabilistic query processing.  Queries processed through Mimir produce
un\-cer\-tain\-ty-annotated result cursors that allow client applications to quickly assess 
result quality and provenance.  We also present a GUI that provides analysts
with an interactive tool for exploring the uncertainty exposed by the system.  Finally, we present optimizations
that make Lenses scalable, and validate this claim through experimental evidence.


\end{abstract}

\keywords{Uncertain Data, Provenance, ETL, Data Cleaning}

\section{Introduction}
\label{sec:introduction}

Data curation is presently performed independently of an analyst's needs.
To trust query results, an analyst first needs to establish trust in her
data, and this process typically requires high upfront investment of human and
computational effort.  However, the level of cleaning effort is often not
commensurate with the specific analysis to be
performed.  A class of so-called
pay-as-you-go~\cite{DBLP:conf/sigmod/JefferyFH08}, or
on-demand~\cite{Yang:2015:LOA:2824032.2824055} data cleaning systems have arisen
to flatten out this upfront cost.  In on-demand cleaning settings, an analyst
quickly applies data cleaning heuristics without needing to tune the process or
supervise the output.  As the analyst poses queries, the on-demand system
continually provides feedback about the quality and precision of the query
results.  If the analyst wishes higher quality, more precise results, the system
can also provide guidance to focus the analyst's data cleaning efforts on
curating inputs that are relevant to the analysis.

In this paper we describe Mimir, a system that extends existing relational 
database engines with support for on-demand curation.  Mimir is based on 
\textbf{lenses}~\cite{Yang:2015:LOA:2824032.2824055}, a powerful and 
flexible new primitive for on-demand curation.  Lenses promise to
enable a new kind of uncertainty-aware data analysis that requires minimal
up-front effort from analysts, without sacrificing trust in the results of that
analysis.  Mimir is presently compatible with SQLite and a popular commercial 
enterprise database management system.

In the work that first introduced Lenses~\cite{Yang:2015:LOA:2824032.2824055} 
we demonstrated how curation tasks including \textit{domain constraint repair}, 
\textit{schema matching}, and \textit{data archival} can be expressed as lenses.
Lenses in general, are a family of unary operators that (1) Apply a data curation heuristic
to clean or validate their input, and
(2) Annotate their output with all assumptions or guesses made by the heuristic.
Critically, lenses require little to no \emph{upfront} configuration --- the
lens' output represents a best-effort guess.  Previous efforts on uncertain data
management~\cite{Fagin:2003:DEG:773153.773163} focus on producing exclusively
correct, or \textit{certain} results.  By comparison, lenses \textit{may} include
incorrect results.  Annotations on the lens output persist through queries and
provide a form of provenance that helps analysts understand potential sources of
error and their impact on query results.  This in turn allows an analyst to
decide whether or not to trust query results, and how to best allocate limited 
resources to data curation efforts.

\begin{figure}
	\newcommand{\tabincell}[2]{\begin{tabular}{@{}#1@{}}#2\end{tabular}}
	\centering
	\tiny
	\begin{tabular}{|c|c|c|c|c|}
		\hline
		\multicolumn{5}{|c|}{Product}\\ \hline
		\textbf{id} & \textbf{name} & \textbf{brand} & \textbf{cat} & \texttt{ROWID} \\
		\hline P123 & Apple 6s, White & \nonDetColumn{?} & phone & R1 \\
		\hline P124 & Apple 5s, Black & \nonDetColumn{?} & phone & R2   \\
		\hline P125 & Samsung Note2 & Samsung & phone  & R3  \\
		\hline P2345 & Sony to inches & \nonDetColumn{?} & \nonDetColumn{?} & R4 \\
		\hline P34234 & Dell, Intel 4 core & Dell & laptop  & R5 \\
		\hline P34235 & HP, AMD 2 core & HP & laptop  & R6 \\ \hline
	\end{tabular}
	\begin{tabular}{|c|c|c|c|c|}
		\hline
		\multicolumn{5}{|c|}{Ratings1}\\ \hline
		\textbf{pid} &{\ldots} &\textbf{rating} &\textbf{review\_ct} & \texttt{ROWID}\\
		\hline P123 & \ldots & 4.5 & 50 &  R7 \\
		\hline P2345 & \ldots & \nonDetColumn{?} & 245 & R8 \\
		\hline P124 & \ldots & 4 & 100 & R9 \\  \hline
	\end{tabular}
	\begin{tabular}{|c|c|c|c|c|c|}
		\hline
		\multicolumn{5}{|c|}{Ratings2}\\ \hline
		\textbf{pid} &{\ldots} &\textbf{evaluation} &\textbf{num\_ratings} & \texttt{ROWID} \\
		\hline P125 & \ldots & 3 & 121 & R10\\
		\hline P34234 & \ldots & 5 & 5 & R11 \\
		\hline P34235 & \ldots & 4.5 & 4 & R12\\ \hline
	\end{tabular}
	\caption {Incomplete error-filled example relations, including an implicit unique identifier attribute \texttt{ROWID}.}
	\label{fig:data}
	\vspace*{-0.1 in}
\end{figure}

\begin{example}
\label{ex:aliceBegins}
Alice is an analyst at a retail store and is developing a 
promotional strategy based on public opinion ratings gathered by two 
data collection companies.  
A thorough analysis of the data requires substantial data curation effort from Alice: 
As shown in Figure~\ref{fig:data}, the rating company's schemas are incompatible, 
and the store's own product data is incomplete.  
However, Alice's preliminary analysis is purely exploratory, and she is hesitant 
to invest the  effort required to fully curate this data.
She creates a lens to fix missing values in the Product table:
{\small
\begin{lstlisting}
CREATE LENS SaneProduct AS SELECT * FROM Product
USING DOMAIN_REPAIR(cat string NOT NULL,
                    brand string NOT NULL);
\end{lstlisting}
}
From Alice's perspective, the lens \texttt{SaneProduct} behaves as a standard database view.
However, the content of the lens is guaranteed to satisfy the domain constraints on \texttt{category} and \texttt{brand}.
\texttt{NULL} values in these columns are replaced according to a classifier built over the \texttt{Product} table. Under the hood, the Mimir system maintains a probabilistic version of the view as a so-called Virtual C-Table (VC-Table). A VC-Table cleanly separates the 
\textbf{existence} of uncertainty (e.g., the category value of a tuple is unknown), the 
\textbf{explanation} for how the uncertainty affects a query result (this is a specific type of provenance), and the \textbf{model} for this uncertainty as a probability distribution (e.g., a classifier for category values that is built when the lens is created).
\end{example}

Uncertainty is encoded in a VC-Table through attribute values that are symbolic expressions over
variables representing unknowns. A probabilistic model for these variables is maintained separately. 
Queries over a VC-Table can be translated into deterministic SQL over the lens' deterministic inputs.
This is achieved by evaluating deterministic expressions as usual and by manipulating symbolic expressions for computations that involve variables. The result is again a relational encoding of a VC-Table. The probabilistic model (or an approximation thereof) can be ``plugged'' into the expressions in a post-processing step to get a deterministic result. This approach has several advantages: (1) the probabilistic model can be created in a pay-as-you-go fashion focusing efforts on the part that is relevant for an analyst's query; (2) the symbolic expressions of a VC-Table serve as a type of provenance that explain how the uncertainty affects the query result; (3) changes to the probabilistic model or switching between different approximations for a model only require repetition of the post-processing step over the already computed symbolic query result; (4) large parts of the computation can be outsourced to a classical relational database; and (5) queries over a mixture of VC-Tables and deterministic tables are supported out of the box.
A limitation of our preliminary work with VC-Tables is the scalability
of the expressions outsourced to the deterministic database.
Our initial approach sometimes creates outsourced expressions
that can not be evaluated efficiently.  
In this paper, we address this limitation, and in doing so demonstrate that 
VC-Tables are a scalable and practical tool for managing uncertain data.
Concretely, in this paper:
\begin{itemize}
\item In Section~\ref{sec:system}, we describe the Mimir system, including its APIs, its user interface, and a novel ``annotated'' result cursor that enables uncertainty-aware analytics.
\item In Section~\ref{sec:optimization}, we demonstrate the limitations of naive VC-Tables and introduce techniques for scalable query processing over VC-Tables.
\item In Section~\ref{sec:experiments}, we evaluate Mimir on SQLite and a commercial database system.
\end{itemize}


\section{Background}
\label{sec:background}
\begin{figure}
\begin{eqnarray*}
e & := & \mathbb R 
	\ordefn Column 
	\ordefn	\ctifthenelse{\phi}{e}{e} \\ &&
	\ordefn e\ \{ +,-,\times,\div \}\ e 
	\ordefn	Var(id[, e[, e[ , \ldots]]]) 
\\
\phi & := & e\ \{ =, \neq, <, \leq, >, \geq \}\ e\ordefn
	\phi\ \{ \wedge, \vee \}\ \phi \ordefn
	\top\ordefn \bot \\ &&
\ordefn
	e\ \texttt{is null}\ordefn
	\neg\phi
\end{eqnarray*}~\\[-0.3in]
\caption{Grammars for boolean expressions $\phi$ and numerical expressions $e$ including VG-Functions $Var(\ldots)$.}
\label{fig:exprgram}
\end{figure} 

\tinysection{Possible Worlds Semantics}
An uncertain database $\mathcal D$ 
over a schema $sch(\mathcal D)$ is defined as a set of possible worlds: 
deterministic database instances $D \in \mathcal D$ over schema 
$sch(D) = sch(\mathcal D)$.  Possible worlds semantics defines queries over 
uncertain databases in terms of deterministic query semantics.  A deterministic 
query $Q$ applied to an uncertain database defines a set of possible results 
$Q(\mathcal D) = \{\;Q(D)\;|\;D \in \mathcal D\;\}$.  Note that these semantics 
are agnostic to the data 
representation, query language, and number of possible worlds $|\mathcal D|$. 
A {\em probabilistic database} 
$\tuple{\mathcal D, p}$ is an uncertain database 
annotated with a probability distribution $p : \mathcal D \rightarrow [0,1]$ that induces a distribution 
over all possible result relations $R$:
$$P[Q(\mathcal D) = R] = \sum_{D \in \mathcal D\;:\;Q(D) = R} p(D)$$


A probabilistic query processing (PQP) system is supposed to answer a 
deterministic query $Q$ by listing all its possible answers and annotating each 
tuple with its marginal probability. These tasks are often \#P-hard in practice, 
necessitating the use of approximation techniques.

\tinysection{C-Tables and PC-Tables}
One way to make probabilistic query processing efficient is to encode $\D$ and 
$P$ through a compact, factorized representation. In this paper we adopt a 
generalized form of C-Tables~\cite{DBLP:journals/jacm/ImielinskiL84,
kennedy2010pip} to represent $\D$, and PC-Tables~\cite{
DBLP:journals/debu/GreenT06,Green:2012} to represent the pair
$(\D, P)$. A C-Table \cite{DBLP:journals/jacm/ImielinskiL84} is a relation 
instance where each tuple is annotated with a formula $\phi$, a 
propositional formula over an alphabet of variable symbols $\Sigma$. The formula
$\phi$ is often called a \emph{local condition} and the symbols in $\Sigma$ are 
referred to as \emph{labeled nulls}, or just variables.
Intuitively, for each assignment to the variables in $\Sigma$ we obtain a 
possible relation containing all the tuples whose formula $\phi$ is satisfied. 
For example:

\begin{center}
  \begin{minipage}{1\linewidth}
    \begin{minipage}{0.79\linewidth}
      \hspace{-3mm}
      {\tiny
        \begin{tabular}{r|c|c|c|c|c|c|}
          \cline{2-6}
          & \multicolumn{5}{|c|}{\bf Product}\\ \cline{2-6}
          & \textbf{pid} & \textbf{name} & \textbf{brand} & \textbf{category} & $\phi$ \\
          \cline{2-6} $t_{1}$ & P123 &  Apple 6s & Apple & phone  & $x_{1}=1$ \\
          \cline{2-6} $t_{2}$ & P123 &  Apple 6s & Cupertino & phone  & $x_{1}=2$ \\
          \cline{2-6} $t_{3}$ & P125 &  Note2 & Samsung & phone  & $\top$  \\
          \cline{2-6} 
        \end{tabular}
      }
    \end{minipage}
    \begin{minipage}{0.18\linewidth}
      {\tiny
        \begin{align*}
          x_1 = 
          \begin{cases}
            1: 0.3\\
            2: 0.7\\
          \end{cases}
        \end{align*}
      }
    \end{minipage}
  \end{minipage}
\end{center}
The above C-Table defines a set of two possible worlds, $\{t_{1}, t_{3}\}$, $\{t_{2}, t_{3}\}$, i.e. one world for each possible assignment to the variables in the one-symbol alphabet $\Sigma=\{x_{1}\}$. Notice that no possible world can have both $t_{1}$ and $t_{2}$ at the same time.
Adding a probabilistic model for the variables, e.g., $P(x_1)$ as shown above, we get a PC-table. For instance, in this example the probability that the brand of product $P123$ is \textit{Apple} is $0.3$. 
 C-Tables are closed w.r.t. positive relational algebra \cite{DBLP:journals/jacm/ImielinskiL84} : if $\D$ is representable by a C-Table and $Q$ is a positive query then $\D'= \comprehension{Q(D)}{D\in\D}$ is representable by another C-Table. 

\tinysection{VG-Relational Algebra}
VG-RA (variable-generating relational algebra)~\cite{kennedy2010pip} is a 
generalization of positive bag-relation algebra with extended projection, that 
uses a simplified form of VG-functions~\cite{jampani2008mcdb}. 
In VG-RA, VG-functions (i) dynamically introduce new Skolem symbols in $\Sigma$,
that are guaranteed to be unique and deterministically derived by the function's
parameters, and (ii) associate the new symbols with probability distributions. 
Hence, VG-RA can be used to define new PC-Tables. 
Primitive-valued expressions in VG-RA (i.e., projection expressions and 
selection predicates) use the grammar summarized in Figure~\ref{fig:exprgram}.  
The primary addition of this grammar is the VG-Function term representing unknown values: $Var(\ldots)$.

VG-RA's expression language enables a generalized form of C-Tables, where 
attribute-level uncertainty is encoded by replacing missing values with VG-RA 
\emph{expressions} (not just variables) that act as freshly defined Skolem 
terms.  For example, the previous PC-Table is equivalent to the generalized
PC-Table:   \\[-10mm]
\begin{center}
  \begin{minipage}{1.05\linewidth}
    \begin{minipage}{0.6\linewidth}
      {\tiny
        \begin{tabular}{|c|c|c|c|c|}
          \hline
          \multicolumn{4}{|c|}{Product}\\ \hline
          \textbf{pid} & \textbf{name} & \textbf{brand} & \textbf{category}  \\
          \hline P123 &  Apple 6s & \nonDetColumn{$Var(\textrm{\textquotesingle} X \textrm{\textquotesingle}, \texttt{R1})$} & phone  \\
          \hline P125 &  Note2 & Samsung & phone \\
          \hline 
        \end{tabular}
      }
    \end{minipage}
    \hspace*{-2mm}
\begin{minipage}{0.4\linewidth}
      {\tiny
        \begin{multline*}
          Var(\textrm{\textquotesingle}X\textrm{\textquotesingle}, \texttt{R1}) \\
          = 
          \begin{cases}
            Apple: 0.3\\
            Cupertino: 0.7\\
          \end{cases}
        \end{multline*}
      }
    \end{minipage}
  \end{minipage}
\end{center}~\\[-5mm]
\indent It has been shown that generalized C-Tables are closed w.r.t VG-RA~\cite{DBLP:journals/jacm/ImielinskiL84,kennedy2010pip}. 
Evaluation rules for VG-RA use a \textit{lazy evaluation} operator
$\evallazy{\cdot}$, which uses a \textit{partial} binding of $Column$ or 
$Var(\ldots)$ atoms to corresponding expressions.
Lazy evaluation applies the partial binding and then reduces every sub-tree in 
the expression that can be deterministically evaluated.  Non-deterministic 
sub-trees are left intact.

Any tuple attribute appearing in a C-Table can be encoded as an abstract syntax 
tree for a partially evaluated expression that assigns it a value.
This is the basis for evaluating projection operators, where every expression 
$e_i$ in the projection's target list is lazily evaluated.  Column bindings are 
given by each tuple in the source relation.
The local condition $\phi$ is preserved intact through the projection.  
Selection is evaluated by combining the selection predicate $\phi$ with each 
tuple's existing local condition.
For example, consider a query $\pi_{brand, category}(\sigma_{brand=Apple}(Product))$ over the example PC-Table. The result of this query is shown below. The second tuple of the input table does not fulfil the selection condition and is thus guaranteed to  not be in the result. Note the symbolic expressions in the local condition and attribute values. Furthermore, note that the probabilistic model for the single variable is not influenced by the query at all.

\begin{center}
  \begin{minipage}{1\linewidth}
    \begin{minipage}{1\linewidth}
      \centering
      {\tiny
        \begin{tabular}{|c|c|c|}
          \hline
          \multicolumn{3}{|c|}{Query Result} \\ \hline
          \textbf{brand} & \textbf{category} & $\phi$  \\
          \hline \nonDetColumn{$Var(\textrm{\textquotesingle} X \textrm{\textquotesingle}, \texttt{R1})$} & phone 
                         &  \nonDetColumn{$Var(\textrm{\textquotesingle} X \textrm{\textquotesingle}, \texttt{R1}) = Apple$}\\
          \hline 
        \end{tabular}
      }
    \end{minipage}
%
  \end{minipage}
\end{center} 
From now on, we will implicitly assume this generalized form of C-Tables.
  
\tinysection{Lenses}
Lenses use VG-RA queries to define new C-Tables as views: A lens defines an 
uncertain view relation through a VG-RA query  $\F_{lens}(Q(D))$, where $\F$ and $Q$ to represents the non-deterministic an deterministic components of 
the query, respectively.
Independently, the lens constructs $P$ as a joint probability distribution over 
every variable introduced by $\F_{lens}$, by defining a sampling process in the 
style of classical VG-functions~\cite{jampani2008mcdb}, or supplementing it with
additional meta-data to create a PIP-style grey-box~\cite{kennedy2010pip}. 
These semantics are closed over PC-Tables.  If $Q(D)$ is non-deterministic --- 
that is, the lens' input is defined by a PC-Table $(Q(D), P_Q)$ --- the lens' 
semantics are virtually unchanged due to the closure of VG-RA over C-Tables. 

\begin{example}
\label{ex:lensDefn}
Recall the lens definition from Example~\ref{ex:aliceBegins}.  This lens defines
a new C-Table using the VG-RA query:
$$\pi_{id \leftarrow id, name \leftarrow name, brand \leftarrow f(brand), cat \leftarrow f(cat)}(Product)$$
In this expression $f$ denotes a check for domain compliance, and a replacement 
with a non-deterministic value if the check fails, as follows:
{\em $$f(x) \equiv \textbf{if}\;x\;\texttt{is null}\;\textbf{then}\;Var(x, \texttt{ROWID})\;\textbf{else}\;x$$}
The models for $Var('brand', \texttt{ROWID})$ and $Var('cat', \texttt{ROWID})$ are defined by  
classifiers trained on the contents of $Product$.
\end{example}

\tinysection{Virtual C-Tables}
Consider a probabilistic database in which all non-determinism is derived from 
lenses.
In this database, all C-Tables, including those resulting from deterministic 
queries over non-deterministic data can be expressed as VG-RA queries over
a deterministic database $D$.
Furthermore, VG-RA admits a normal form~\cite{Yang:2015:LOA:2824032.2824055} 
for queries where queries are segmented into a purely deterministic component 
$Q(D)$ and a non-deterministic component $\F(Q(D))$.  These normalization rules 
are shown in Figure~\ref{fig:normalFormReduction}.

\begin{figure*}
\centering
\begin{eqnarray}
\label{rewrite:projection}
\pi_{a_j' \leftarrow e_j'} \left(\F(\tuple{a_i \leftarrow e_i}, \phi)(Q(D))\right) 
 \equiv \F(\tuple{a_j' \leftarrow \evallazy{e_j'(a_i \leftarrow e_i)}}, \phi)(Q(D))
\end{eqnarray}\vspace*{-0.2in}
\begin{eqnarray}
\label{rewrite:selection}
\sigma_{\psi} \left(\F(\tuple{a_i \leftarrow e_i}, \phi)(Q(D))\right)
 \equiv \F(\tuple{a_i \leftarrow e_i}, \phi \wedge \psi_{var})(\sigma_{\psi_{det}}(Q(D)))
\end{eqnarray}\vspace*{-0.2in}
\begin{eqnarray}
\label{rewrite:cross}
\F(\tuple{a_i \leftarrow e_i}, \phi)(Q(D)) \times \F(\tuple{a_j' \leftarrow e_j'}, \phi')(Q'(D))
  \equiv \F(\tuple{a_i \leftarrow e_i, a_j' \leftarrow e_j'}, \phi \wedge \phi')(Q(D) \times Q'(D))
\end{eqnarray}\vspace*{-0.2in}
\begin{multline}
\label{rewrite:union}
\F(\tuple{a_i \leftarrow e_i}, \phi)(Q(D)) \uplus \F(\tuple{a_i \leftarrow e_i'}, \phi')(Q'(D))
  \equiv\\ \F(\tuple{a_i \leftarrow
    \evallazy{\ctifthenelse{src=1}{e_i}{e_i'}}}, 
    \evallazy{\ctifthenelse{src=1}{\phi}{\phi'}}
  )(
  \pi_{*, src \leftarrow 1}(Q(D)) \uplus \pi_{*, src \leftarrow 2}(Q'(D)))
\end{multline}
\caption{Reduction to VG-RA Normal Form.}
\label{fig:normalFormReduction}
\end{figure*}

Normalization does not affect the linkage between the C-Table 
computed by a VG-RA query and its associated probability measure $P$: $Var(\ldots)$ 
remains unchanged.  Moreover, the non-deterministic component of the normal
form $\F$ is a simple composite projection and selection operation.

The simplicity of $\F$ carries two benefits.  First, the deterministic component
of the query can be evaluated natively in a database engine, while the 
non-deterministic component can be applied through a simple shim
interface wrapping around the database.  Second, the abstract syntax tree of the
expression acts a form of provenance~\cite{arab2014generic,arab2014reenacting} 
that annotates uncertain query results with metadata describing the level and
nature of their uncertainty, a key component of the system we now describe. For example, in the query result shown above it is evident that the tuple will be in the result as long as the condition \nonDetColumn{$Var(\textrm{\textquotesingle} X \textrm{\textquotesingle}$, \texttt{R1}) = `Apple'} evaluates to true. Mimir provides an API for the user to retrieve this type of explanation for a query result and comes with a user interface that visualizes explanations.


\section{System Outline}
\label{sec:system}

The Mimir system is a shim layer that wraps around an existing DBMS to provide
support for lenses.
Using Mimir, users define lenses that perform common data cleaning operations
such as schema matching, missing value interpolation, or type inference with little
or no configuration on the user's part.
Mimir exports a native SQL query interface that allows lenses to be queried 
as if they were ordinary relations in the backend database.
A key design feature of Mimir is that it has minimal impact on its environment.
Apart from using the backend database to persist metadata, Mimir does not modify 
the database or its data in any way.  
As a consequence, Mimir can be used alongside any existing database workflow with 
minimal effort and minimal risk.

\subsection{User Interface}
Users define lenses through a \texttt{CREATE LENS} statement that immediately 
instantiates a new lens.  
\begin{example}
Recall the example data from Figure~\ref{fig:data}.  To merge the two ratings
relations, Alice needs to re-map the attributes of \texttt{Ratings2}.  Rather
than doing so manually, she defines a lens that re-maps 
the attributes of the \texttt{Ratings2} relation to those of \texttt{Ratings1}
as follows.  
\pagebreak
\begin{lstlisting}
CREATE LENS MatchedRatings2 AS
SELECT * FROM Ratings2 
USING SCHEMA_MATCHING(pid string, ..., 
      rating float, review_ct float, NO LIMIT);
\end{lstlisting}

\texttt{CREATE LENS} statements behave like a view definition, but also apply 
a data curation step to the output; in this case schema matching.
Mapping targets may be defined explicitly or by selecting an 
existing relation's schema in the GUI.
\end{example}

\begin{figure}
\centering
\includegraphics[width=\columnwidth]{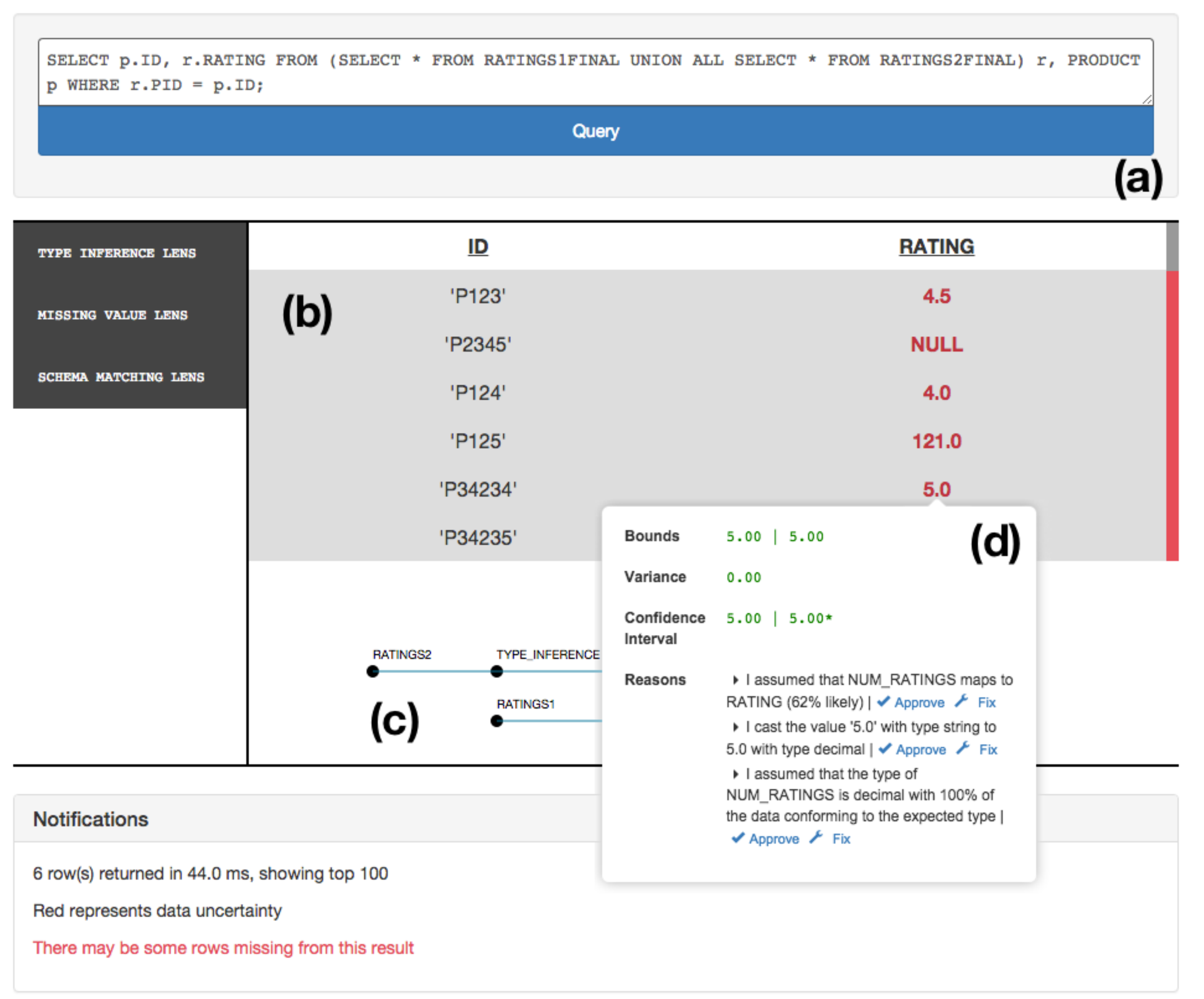}
\caption{The Graphical Mimir User Interface}
\label{fig:ui}
\end{figure}

In addition to a command-line tool, Mimir provides a Graphical User Interface (GUI) illustrated in Figure~\ref{fig:ui}.
Users pose queries over lenses and deterministic relations using standard SQL 
\texttt{SELECT} statements (a).  
Mimir responds to queries over lenses with a \textit{best guess} result, or
the result of the query in the possible world with maximum likelihood.  
In contrast to the classical notion of ``certain'' answers, the best guess 
\textit{may} contain inaccuracies.  However, all uncertainty arises from 
$Var$ terms introduced by lenses.  Consequently, using the provenance
of each row and cell, Mimir can identify potential sources of error: Attribute 
values that depend on a $Var$ term may be incorrect, and filtering 
predicates that depend on a $Var$ term may lead to rows incorrectly 
being included or excluded in the result.  We refer to these two types of
error as \textit{non-deterministic cells}, and \textit{non-deterministic rows}, respectively.
\begin{example}
Recall the result of the example query in Section~\ref{sec:background}, which shows a VC-table before the best guess values are plugged in. The only row is non-deterministic, because its existence depends on the value of \nonDetColumn{$Var(\textrm{\textquotesingle} X \textrm{\textquotesingle}, \texttt{R1})$} which denotes the unknown brand of this tuple. The brand attribute value of this tuple is a non-deterministic cell, because its value depends on the same expression.
\end{example}

In Mimir, query results (b) visually convey potential sources of error through several 
simple cues.  
First, a small provenance graph (c) helps the user quickly identify the data's origin, 
what cleaning heuristics have been applied, and where.

Potentially erroneous results are clearly identified: Non-deterministic
rows have a red marker on the right and a grey background, while non-deterministic
cells are highlighted in red.

Clicking on a non-deterministic row or cell brings up an explanation window (d).  
Here, Mimir provides the user with statistical metrics summarizing the 
uncertainty of the result, as well as a list of human-readable reasons why the 
result might be incorrect.  
Each reason is linked to a specific lens; If the user believes a reason to be 
incorrect, she can click ``Fix'' to override the lens' data cleaning decision.
An ``Approve'' button allows a user to indicate that the lens heuristic's 
choice is satisfactory.  
Once all reasons for a given row or value's non-determinism have been either 
approved or fixed, the row or value becomes green to signify that it is 
now deterministic.

\begin{example}
Figure~\ref{fig:ui} shows the results of a query where one product (with id `P125') has an 
unusually high rating of 121.0.  
By clicking on it, Alice finds that a schema matching lens has incorrectly mapped 
the \texttt{\upshape NUM\_RATINGS} column of one input relation to the \texttt{\upshape RATINGS} 
column of the other input relation --- 121 is the number of ratings for the product, not the 
actual rating itself.  
By clicking on the fix button, Alice can manually specify the correct match
and Mimir re-runs the query with the correct mapping.
\end{example}

Making sources of uncertainty easily accessible allows the user to quickly 
track down errors that arise during heuristic data cleaning, even while 
viewing the results of complex queries.  
Limiting Mimir to simple signifiers like highlighting and notifications prevents the user 
from being overwhelmed by details, while explanation windows still allow the user to explore 
uncertainty sources in more depth at their own pace.

\subsection{The Mimir API}
\label{sec:api}

\begin{figure}
\centering
\includegraphics[width=0.55\columnwidth]{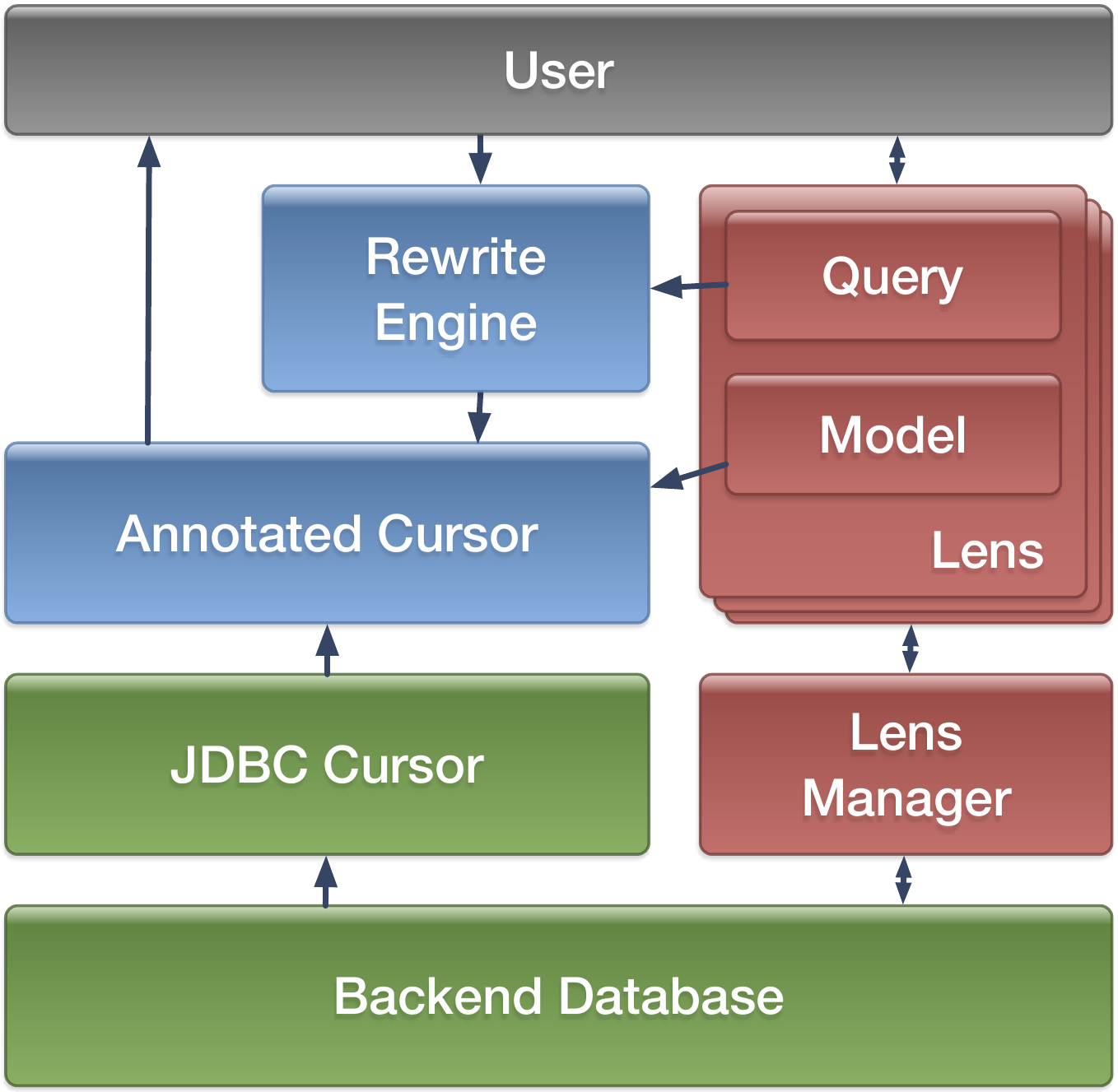}
\caption{The Mimir System}
\label{fig:system}
\end{figure}

The Mimir system's architecture is shown in Figure~\ref{fig:system}.  
 Mimir acts as an intermediary between users and a backend
database using JDBC.  
Mimir exposes the database's native SQL interface, and extends it with support
for lenses.
The central feature of this support is five new functions in the JDBC result 
cursor class that permit client applications such as the Mimir GUI to evaluate result quality.
The first three indicate the presence of specific classes of uncertainty:
(1) \texttt{isColumnDeterministic(int | String)} returns a boolean that 
indicates whether the value of the indicated attribute was computed deterministically without having to ``plug in'' values for variables. 
In our graphical interface, cells for which this function returns false are 
highlighted in red.
Note that the same column may contain both deterministic and non-deterministic
values (e.g., for a lens that replaces missing values with interpolated 
estimates)
(2) \texttt{isRowDeterministic()} returns a boolean that indicates whether
the current row's presence in the output can be determined without using the probabilistic model.
In our graphical interface, rows for which this function returns false are
also highlighted.
(3) \texttt{nonDeterministicRowsMissing()} returns a count of the number
of rows that have been so far ommitted from the result, and were discarded 
based on the output of a lens.  In our graphical interface, when this method 
returns a number greater than zero after the cursor is exhausted, a 
notification is shown on the screen.

As we discuss below, limiting the response of these functions to a simple 
boolean makes it possible to evaluate them rapidly, in-line with the query 
itself.  For additional feedback, Mimir provides two methods: 
\texttt{explainColumn(int | String)} and \texttt{explainRow()}
Both methods construct and return an explanation object as detailed below.
In the graphical Mimir interface, these methods are invoked when a user 
clicks on a non-deterministic (i.e., highlighted) row or cell, and the resulting
explanation object is used to construct the uncertainty summary shown in 
the explanation window.  Explanations do not need to be computed in-line
with the rest of the query, but to maintain user engagement, explanations 
for individual rows or cells still need to be computed quickly when requested.

\subsection{Lens Models}
A lens consists of two components: (1) A VG-RA expression that computes the 
output of the lens, introducing new variables in the process using 
$Var$ terms, and (2) A model object that defines a probability space 
for every introduced variable.  
Recall that $Var$ terms act as skolem functions, introducing new variable 
symbols based on their arguments.  
For example the \texttt{ROWID} attribute can be used to create a distinct variable named ``X'' for every row using the expression $Var('X', \texttt{ROWID})$.
Correspondingly, we distinguish between $Var$ terms and the variable instances 
they create.  Note that the latter is uniquely identified by the name
and arguments of the $Var$ term.
The model object has three mandatory methods:
(1) \texttt{getBestGuess(var)} returns the value of the specified variable instance in the most likely possible world.
(2) \texttt{getSample(var, id)} returns the value of the specified variable 
instance in a randomly selected possible world.  $id$ acts
as a seed value, ensuring that the same possible world is selected across 
multiple calls.
(3) \texttt{getReason(var)} returns a human-readable explanation of the
heuristic guess represented by the specified variable instance.
The \texttt{getBestGuess} method is used to produce best-guess query results.
The remaining two methods are used by explanation objects.  As in
PIP~\cite{kennedy2010pip} and Orion 2.0~\cite{Singh:2008:ONS:1376616.1376744}, optional metadata can sometimes permit the use of closed-form solutions
when computing statistical metrics for result values.

\subsection{Explanation Objects}
Explanation objects provide a means for client applications like the GUI to 
programmatically analyze the non-determinism of a specific row or cell.  
Concretely, an explanation object provides methods that compute: 
(1) Statistical metrics that quantitatively summarize the distribution of possible result outcomes, 
and (2) Qualitative summaries or depictions of non-determinism in the result.

\subsubsection{Statistical Metrics} 
Available statistical metrics depend on whether the explanation object is
constructed for a row or a cell, and in the latter case also on what type of
value is contained in the cell.  
For rows, the explanation object has only one method that computes
the row's confidence, or the probability that the row is part of the result set.

For numerical cells, the explanation object has methods for computing
the value's variance, confidence intervals, and may also be able to provide
upper and lower bounds.
Variance and confidence intervals are computed analytically if possible, or
Monte Carlo style by generating samples of the result using the 
\texttt{getSample} method on all involved models.  
When computing these metrics using Monte Carlo, we discard samples
that do not satisfy the local condition of the cell's row, as the cell will
not appear in the result at all if the local condition is false.
Statistical metrics are not computed for non-numerical cells.

\subsubsection{Qualitative Summaries}
Quantitative statistical metrics do not always provide the correct intuition
about a result value's quality.  In addition to the above metrics, an 
explanation object can also use Monte Carlo sampling to construct 
histograms and example result values for non-deterministic cells.  

Furthermore, for both cells and rows, an explanation object can produce
a list of \textit{reasons} --- the human readable summaries obtained from
each participating model's \texttt{getReason} method.  Reasons are 
ranked according to the relative contribution of each $Var$ term to the
uncertainty of the result using a heuristic called
CPI~\cite{Yang:2015:LOA:2824032.2824055}.

\subsection{Query Processing}
\label{sec:queryprocessing}
Queries issued to Mimir are parsed into an intermediate representation (IR) 
based on VG-RA.
Mimir maintains a list of all active lenses as a relation in the backend 
database.  
References to lenses in the IR are replaced with the VG-RA expression that 
defines the lens' contents.  
The resulting expression is a VG-RA expression over deterministic data 
residing in the backend database.

Before queries over lenses are evaluated, the query's VG-RA expression is first 
normalized into the form $\F(Q(D))$, where 
$\F(R) = \pi_{a_i\leftarrow e_i}(\sigma_\phi(R))$, 
 $\phi$ represents a non-de\-ter\-min\-is\-tic boolean expression, and the
subsequent projection assigns the result of non-deterministic expressions $e_i$
to the corresponding attributes $a_i$.  
Mimir obtains a classical JDBC cursor for $Q(D)$ from the backend database
and constructs an extended cursor using $\F$ and the JDBC cursor.

Recall  non-determinism in $\phi$ and $e_i$ arises from $Var$ terms.
When evaluating these expressions, Mimir obtains a specific value for the term 
using the \texttt{getBestGuess} method on the model object associated with 
each term.  Apart from this, $\phi$ and the $e_i$ expressions are evaluated as normal.

Determining whether an expression is deterministic or not requires slightly 
more effort.  
In principle, we could say that any expression is non-deterministic if it 
contains a $Var$ term.  
However, it is still possible for the expression's result to be entirely 
agnostic to the value of the $Var$.  
\begin{example}
\label{ex:conditionalNonDeterminism}
Consider the following expression, which is used in Mimir's domain constraint
repair lens:
{\em $$\ctifthenelse{\isnull{A}}{Var('A', \ROWID)}{A}$$}
Here, the value of the expression is only non-deterministic for rows where
$A$ is null.  
\end{example}
Concretely, there are three such cases: (1) conditional expressions where 
the condition is deterministic, (2) \texttt{AND} expressions where one clause
is deterministically false, and (3) \texttt{OR} expressions where one clause 
is deterministically true.  Observe that these cases mirror semantics for 
\texttt{NULL} and \texttt{UNKNOWN} values in deterministic SQL.

For each $e_i$ and $\phi$, the Mimir compiler uses a recursive descent
through the expression illustrated in Algorithm~\ref{alg:isdet} to obtain a 
boolean formula that determines whether the 
expression is deterministic for the current row.  These formulas permit quick
responses to the \texttt{isColumnDeterministic} and 
\texttt{isRowDeterministic} methods.  The counter for the 
\texttt{nonDeterministicRowsMissing} method is computed by using 
\texttt{isRowDeterministic} on each discarded row.

\begin{algorithm}
\caption{\texttt{isDet}($E$)}
\label{alg:isdet}
\begin{algorithmic}[1]
\REQUIRE{$E$: An expression in either grammar from Fig.~\ref{fig:exprgram}.}
\ENSURE{An expression that is true when $E$ is deterministic.}
\IF{$E \in \{\mathbb R, \top, \bot\}$}
\RETURN $\top$
\ELSIF{$E$ \textbf{is} $Var$}
\RETURN $\bot$
\ELSIF{$E$ \textbf{is} $Column_i$}
\RETURN $\top$
\ELSIF{$E$ \textbf{is} $\neg E_1$}
\RETURN $\texttt{isDet}(E_1)$
\ELSIF{$E$ \textbf{is} $E_1 \vee E_2$}
\RETURN $(E_1 \wedge \texttt{isDet}(E_1)) \vee (E_2 \wedge \texttt{isDet}(E_2))$
\STATE \hspace*{20mm} $ \vee~(\texttt{isDet}(E_1) \wedge \texttt{isDet}(E_2))$
\ELSIF{$E$ \textbf{is} $E_1 \wedge E_2$}
\RETURN $(\neg E_1 \wedge \texttt{isDet}(E_1)) \vee (\neg E_2 \wedge \texttt{isDet}(E_2))$
\STATE \hspace*{20mm} $ \vee~(\texttt{isDet}(E_1) \wedge \texttt{isDet}(E_2))$
\ELSIF{$E$ \textbf{is} $E_1\;\{+,-,\times,\div,=,\neq,>,\geq,<,\leq\}\;E_2$}
\RETURN $(\texttt{isDet}(E_1) \wedge \texttt{isDet}(E_2))$
\ELSIF{$E$ \textbf{is} \ctifthenelse{$E_1$}{$E_2$}{$E_3$}}
\RETURN $\texttt{isDet}(E_1) \wedge \left(~~~(E_1 \wedge \texttt{isDet}(E_2))\right.$
\STATE \hspace*{31.75mm} $\left.\vee (\neg E_1 \wedge \texttt{isDet}(E_3))~~\right)$
\ENDIF
\end{algorithmic}
\end{algorithm}

When one of the  explain methods is called, Mimir extracts all
of the $Var$ terms from the corresponding expression, and uses the associated
model object's \texttt{getReason} method to obtain a list of reasons.  
Variance, confidence bounds, and row-level confidence are computed by sampling 
from the possible worlds of the model using \texttt{getSample} and evaluating
the expression in each possible world.  Upper and lower bounds are obtained if
possible from an optional method on the model object, and propagated through 
expressions where possible.


\section{Optimizing Virtual C-Tables}
\label{sec:optimization}

The primary scalability challenge that we address in this paper relates to how queries are normalized in Virtual C-Tables.  Concretely, the problem arises in the rule for normalizing selection predicates:
\begin{multline*}
\sigma_{\psi} \left(\F(\tuple{a_i \leftarrow e_i}, \phi)(Q(D))\right)\\
 \equiv \F(\tuple{a_i \leftarrow e_i}, \phi \wedge \psi_{var})(\sigma_{\psi_{det}}(Q(D)))
\end{multline*}
Non-deterministic predicates are always pushed into $\F$, including those that 
could otherwise be used as join predicates.  When this happens, the backend
database is given a cross-product query to evaluate, and the join is evaluated 
far less efficiently as a selection predicate in the Mimir shim layer.

In this section, we explore variations on the theme of query normalization.  
These alternative evaluation strategies make it possible for a traditional database 
to scalably evaluate
C-Table queries, while retaining the functionality of Mimir's uncertainty-annotated
cursors as described in Section~\ref{sec:api}.  Supporting C-Tables and annotated
cursors carries several challenges:

\tinysection{Var Terms}  
Classical databases are not capable of managing non-determinism, making 
$Var$ terms functionally into black-boxes.  Although a single best-guess value 
does exist for each term, the models that compute this value normally reside 
outside of the database.

\tinysection{\texttt{isDeterministic} methods}
Mimir's annotated cursors must be able to determine whether a given row's 
presence (resp., a cell's value) depends on any $Var$ terms.  Using $\F$, this 
is trivial, as all $Var$ terms are conveniently located in a single expression 
that is used to determine the row's presence (resp., to compute a cell's value).  
Because these methods are used to construct the initial response
shown to the user (i.e., to determine highlighting), they must be fast.

\tinysection{Potentially Missing Rows}
Annotated cursors must also be able to evaluate the number of rows that
could potentially be missing, depending on how the non-determinism is resolved.
Although the result of this method is presented to the user as part of the initial 
query, the value is shown in a notification box and is off of the critical path of 
displaying the best guess results themselves.

\tinysection{Explanations}
The final feature that annotated cursors are expected to support is the
creation of explanation objects.  These do not need to be created until explicitly
requested by the user; the initial database query does not need to be directly
involved in their construction.  However, it must still be possible to construct and
return an explanation object quickly to maintain user engagement.

\medskip

We now discuss two complimentary techniques for constructing annotated 
iterators over Virtual C-Tables.  Our first approach \textit{partitions} queries
into deterministic and non-deterministic fragments to be evaluated 
separately.  The second approach pre-materializes best-guess values into
the backend database, allowing it to evaluate the non-de\-ter\-mi\-nis\-tic query
with $Var$ terms \textit{inlined}.

\subsection{Approach 1: Partition}
\label{sec:partition}

We observe that uncertain data is frequently the minority of the raw data.  Moreover, for some lenses, whether a row is deterministic or not is data-dependent.  
Our first approach makes better use of the backend database by partitioning the query into one or more deterministic and non-deterministic segments, computing each independently, and unioning the results together.  
When the row-determinism $\phi$ of a result depends on deterministic data we can push more work into the backend database for those rows that we know to be deterministic.  For this deterministic partition of the data, joins can be evaluated correctly and other selection predicates can be satisfied using indexes over the base data.
As a further benefit, tuples in each partition share a common lineage, allowing substantial re-use of annotated cursor metadata for all tuples returned by the query on a single partition.
To partition a query $\F(Q(D))$, we begin with a set of partitions, each defined by a boolean formula $\psi_i$ over attributes in $sch(Q)$.  The set of partitions must be complete ($\bigvee \psi_i \equiv \top$) and disjoint ($\forall i\neq j\;.\;\psi_i \rightarrow \neg \psi_j$).  In general, partition formulas are selected such that $\sigma_{\psi_i}(Q(D))$ never contains query results that can be deterministically excluded from $\F(Q(D))$.

\begin{example}
Recall the \texttt{SaneProduct} lens from Examples~\ref{ex:aliceBegins} 
and \ref{ex:lensDefn}.  Alice the analyst now posses a query:
\begin{lstlisting}
SELECT name FROM SaneProduct 
WHERE brand = 'Apple' AND cat = 'phone'
\end{lstlisting}
Some rows of the resulting relation are non-deterministic, but only when the
\texttt{brand} or \texttt{cat} in the corresponding row of \texttt{Product} is \texttt{NULL}.  
Optimizing further, all products that are known to be either non-phones or non-Apple products are also deterministically not in the result.
\end{example}

Given a set of partitions $\Psi = \{\psi_1,\ldots,\psi_N\}$, the partition rewrite transforms the original query into an equivalent set of partitioned queries as follows:
\begin{multline*}
\left(\F(\tuple{a_i \leftarrow e_i}, \phi)(Q(D))\right)\\
\mapsto \F(\tuple{a_i \leftarrow e_i}, \phi_{var,1})(\sigma_{\psi_{1}\wedge\phi_{det,1}}(Q(D)))
\\
\cup\cdots\cup\F(\tuple{a_i \leftarrow e_i}, \phi_{var,N})(\sigma_{\psi_N\wedge\phi_{det,N}}(Q(D)))
\end{multline*}
where $\phi_{var,i}$ and $\phi_{det,i}$ are respectively the non-deterministic and deterministic clauses of $\phi$ (i.e., $\phi = \phi_{var,i} \wedge \phi_{det,i}$) for each partition.
Partitioning then, consists of two stages: (1) Obtaining a set of potential partitions 
$\Psi$ from the original condition $\phi$, and (2) Segmenting $\phi$ into 
a deterministic filtering predicate and a non-deterministic lineage component.

\begin{algorithm}
\caption{$\texttt{naivePartition}(\phi)$}
\label{alg:getpsi}
\begin{algorithmic}
	\REQUIRE{$\phi$: A non-deterministic boolean expression}
	\ENSURE{$\Psi$: A set of partition conditions $\{\psi_i\}$}
	\STATE $clauses \gets \emptyset$
	\STATE $\Psi \gets \emptyset$
	\FOR{$(\ctifthenelse{condition}{\alpha}{\beta}) \in \texttt{subexps}(\phi)$}
    	\STATE \texttt{/*} Check \texttt{if}s in $\phi$ for candidate partition clauses \texttt{*/}
		\IF{$\texttt{isDet}(condition) \wedge (\texttt{isDet}(\alpha) \neq \texttt{isDet}(\beta))$}
			\STATE $clauses \gets clauses \cup \{condition\}$
		\ENDIF
	\ENDFOR	
	\STATE \texttt{/*} Loop over the power-set of clauses \texttt{*/}
	\FOR{$partition \in 2^{clauses}$} 
		\STATE $\psi_i \gets \top$
	    	\STATE \texttt{/*} Clauses in the partition are true, others are false \texttt{*/}
		\FOR{$clause \in clauses$} 
			\STATE \textbf{if} $clause \in partition$ \textbf{then} $\psi_i \gets \psi_i \wedge clause$ 
			\STATE \hspace*{29.5mm} \textbf{else} $\psi_i \gets \psi_i \wedge \neg clause$ 
		\ENDFOR
		\STATE{$\Psi\gets \Psi \cup \{\psi_i\}$}	
	\ENDFOR
\end{algorithmic}
\end{algorithm} 

\subsubsection{Partitioning the Query}

Algorithm~\ref{alg:getpsi} takes the selection predicate $\phi$ in the shim query $\F_{\tuple{a_i \leftarrow e_i}, \phi}$, and outputs a set of partitions $\Psi = \{\psi_i\}$.  Partitions are formed from the set of all possible truth assignments to a set of candidate clauses.  Candidate clauses are obtained from if statements appearing in $\phi$ that have deterministic conditions, and that branch between deterministic and non-deterministic cases.  For example, the if statement in Example~\ref{ex:lensDefn} branches between deterministic values for non-null attributes, and non-deterministic possible replacements.

\begin{example}
The normal form $\F(Q(D))$ of the query in the prior example has the non-deterministic condition ($\phi$):\vspace*{-2mm}
{\small \em
\begin{multline*}
(\ctifthenelse{\isnull{brand}}{Var('b',\ROWID)}{brand}) = `Apple'\\
\wedge(\ctifthenelse{\isnull{cat}}{Var('c',\ROWID)}{cat}) = `phone'\vspace*{-2mm}
\end{multline*}
}
There are two candidate clauses in $\phi$: $\isnull{brand}$ and $\isnull{cat}$. Thus, Algorithm~\ref{alg:getpsi} creates 4 partitions: 
{\em $\psi_1 =(\neg\isnull{brand}\wedge \neg\isnull{cat})$}, 
{\em $\psi_2 =(    \isnull{brand}\wedge \neg\isnull{cat})$}, 
{\em $\psi_3 =(\neg\isnull{brand}\wedge     \isnull{cat})$}, and finally
{\em $\psi_4 =(    \isnull{brand}\wedge     \isnull{cat})$}.
\end{example}

\subsubsection{Segmenting $\phi$} 

For each partition $\psi_i$ we can simplify $\phi$ into a reduced form $\phi_i$.  We use $\applyBind{\psi_i}{\phi}$ to denote the result of propagating the implications of $\psi_i$ on $\phi$.  For example, $(\textbf{if}\ {\isnull{X}}\ \textbf{then}$ ${Var('X')}\ \textbf{else}\ {X})[\isnull{X}] \equiv Var('X')$.
Using \texttt{isDet} from Algorithm~\ref{alg:isdet}, we partition the conjunctive terms of $\applyBind{\psi_i}{\phi}$ into deterministic and non-deterministic components $\phi_{i,det}$ and $\phi_{i,var}$, respectively so that \\[-5mm]
$$(\phi_{i,det}\wedge\phi_{i,var}) \equiv \applyBind{\psi_i}{\phi}$$

\subsubsection{Partitioning Complex Boolean Formulas} 

As discussed in Section~\ref{sec:queryprocessing} there are three cases where non-determinism can be data-dependent: conditional expressions, conjunctions, and disjunctions.  Algorithm~\ref{alg:getpsi} naively targets only conditionals.  Conjunctions come for free, because deterministic clauses can be freely migrated into the deterministic query already.  However, queries including disjunctions can be further simplified. 

\begin{example}
We return once again to our running example, but this time with a disjunction in the \texttt{WHERE} clause
\begin{lstlisting}
SELECT name FROM SaneProduct 
WHERE brand = 'Apple' OR cat = 'phone'
\end{lstlisting}
Propagating $\psi_3$ into the normalized condition $\phi$ gives:
$$(\applyBind{\psi_3}{\phi}) \equiv \left(brand='Apple' \vee Var('c',\ROWID)='phone'\right)$$
The output is always deterministic for rows where $brand='Apple'$.  However, this formula can not be subdivided into deterministic and non-deterministic components as above.  
\end{example}

We next describe a more aggressive partitioning strategy that uses the structure of $\phi$ to create partitions where each partition depends on exactly the same set of $Var$ terms.
To determine the set of partitions for each sub-query, we use a recursive traversal through the structure of $\phi$, as shown in in Algorithm~\ref{algo:optgetpsi}.  In contrast to the naive partitioning scheme, this algorithm explicitly identifies two partitions where $\phi$ is deterministically true and deterministically false.  This additional information helps to exclude cases where one clause of an OR (resp., AND) is deterministically true (resp., false) from the non-deterministic partitions.  To illustrate, consider the disjunction case handled by Algorithm~\ref{algo:optgetpsi}.  In addition to the partition where both children are non-deterministic, the algorithm explicitly distinguishes two partitions where one child is non-deterministic and the other is deterministically false.  When $\phi$ is segmented, the resulting non-deterministic condition for this partition will be simpler.

\begin{algorithm}
	\caption{$\texttt{generalPartition}(\phi)$}
	\label{algo:optgetpsi}
	\begin{algorithmic}
	\REQUIRE{$\phi$: A non-deterministic boolean expression.}
	\ENSURE{$\psi_{\top}$: The partition where $\phi$ is deterministically true.}
	\ENSURE{$\psi_{\bot}$: The partition where $\phi$ is deterministically false.}
	\ENSURE{$\Psi_{var}$: The \textit{set} of non-deterministic partitions.}
		\IF{$\phi$ \textbf{is} $\phi_1 \vee \phi_2$}
			\STATE $\tuple{\psi_{\top,1}, \psi_{\bot,1}, \Psi_{var,1}} \gets \texttt{generalPartition}(\phi_1)$
			\STATE $\tuple{\psi_{\top,2}, \psi_{\bot,2}, \Psi_{var,2}} \gets \texttt{generalPartition}(\phi_2)$
			\STATE $\psi_{\top} \gets \psi_{\top,1} \vee \psi_{\top,2}$
			\STATE $\psi_{\bot} \gets \psi_{\bot,1} \wedge \psi_{\bot,2}$
			\FORALL{$\psi_{var,1},\psi_{var,2} \in \Psi_{var,1}, \Psi_{var,2}$}
				\STATE $\Psi \gets \Psi \cup \{ \psi_{var,1} \wedge \psi_{var,2} \}$
			\ENDFOR
			\STATE \textbf{for all} $\psi_{var,1} \in \Psi_{var,1}$ \textbf{do} 
				$\Psi \gets \Psi \cup \{ \psi_{var,1} \wedge \psi_{\bot,2} \}$
			\STATE \textbf{for all} $\psi_{var,2} \in \Psi_{var,2}$ \textbf{do} 
				$\Psi \gets \Psi \cup \{ \psi_{var,2} \wedge \psi_{\bot,1} \}$
		\ELSIF{$\phi$ \textbf{is} $\phi_1 \wedge \phi_2$}
			\STATE \ccomment{Symmetric with disjunction}
		\ELSIF{$\phi$ \textbf{is} $\neg \phi_1$}
			\STATE $\tuple{\psi_{\top,1}, \psi_{\bot,1}, \Psi_{var,1}} \gets \texttt{generalPartition}(\phi_1)$
			\STATE $\tuple{\psi_{\bot}, \psi_{\top}, \Psi_{var}} = \tuple{\psi_{\top,1}, \psi_{\bot,1}, \Psi_{var,1}}$
		\ELSE
			\STATE $\Psi = \texttt{naivePartition}(\phi)$
			\STATE $\Psi_{det} \gets \emptyset$; $\Psi_{var} \gets \emptyset$
			\FORALL{$\psi \in \Psi$}
				\STATE \textbf{if} $\texttt{isDet}(\applyBind{\psi}{\phi})$ \textbf{then}
					$\Psi_{det} \gets \Psi_{det} \cup \{ \psi \}$
				\STATE \hspace*{22.5mm} \textbf{else}
					$\Psi_{var} \gets \Psi_{var} \cup \{ \psi \}$
			\ENDFOR
			\STATE $\psi_{\top} = \left(\bigvee \Psi_{det}\right) \wedge \applyBind{\bigvee \Psi_{det}}{\phi} $ 
			\STATE $\psi_{\bot} = \left(\bigvee \Psi_{det}\right) \wedge \applyBind{\bigvee \Psi_{det}}{\neg\phi} $ 
		\ENDIF
	\end{algorithmic}
\end{algorithm}

\medskip

The partition approach makes full use of the backend database engine by splitting the query into deterministic and non-deterministic fragments.  The lineage of the condition for each sub-query is simpler, and generally not data-dependent for all rows in a partition.  As a consequence, explanation objects can be shared across all rows in the partition.
The number of partitions obtained with both partitioning schemes is exponential in the number of candidate clauses.  Partitions could conceivably be combined, increasing the number of redundant tuples processed by Mimir to create a lower-complexity query.  In the extreme, we might have only two partitions: one deterministic and one non-deterministic.  We leave the design of such a partition optimizer to future work.


\subsection{Approach 2: Inline}
\label{sec:inline}
During best-guess query evaluation, each variable instance is replaced by a 
single, deterministic best-guess.  Simply put, best-guess queries are themselves 
deterministic.  The second approach exploits this observation to directly offload 
virtually all computation into the database.
Best-guess values for all variable instances are 
pre-materialized into the database, and the $Var$ terms themselves are
replaced by nested lookup queries that can be evaluated directly.

\subsubsection{Best-Guess Materialization}

As part of lens creation, best-guess estimates must now be materialized. Recall from the grammar in Figure~\ref{fig:exprgram}, $Var(id, e_1, \ldots, e_n)$ terms are defined by a unique identifier id and zero or more parameters ($e_i$). 
For each unique variable identifier allocated by the lens, Mimir creates a new table in the database.
The schema of the best-guess table consists of the variable's parameters ($e_i$), a best guess value for the variable, and other metadata for the variable including whether the user ``Accept''ed it. 
The variable parameters form a key for the best-guess table.

\begin{example}
Recall the domain repair lens from Example~\ref{ex:lensDefn}. To materialize the best-guess relation, Mimir run the lens query to determine all variable instances that are used in the current database instance.  In the example, there are 4 such variables, one for each null value. For instance, the missing brand of product $P123$ will instantiate a variable $Var('brand','P123')$. For all ``brand'' variables Mimir will create a best guesses table with primary key \texttt{param1}, a best-guess value \texttt{value}, and attributes storing the additional metadata mentioned above. Mentions of variables in queries over the lens are replaced with a subquery that returns the best guess value. For example, in $\projection_{brand}(SaneProduct)$ the expression for \texttt{brand} in the VG-RA query:
{\em $$\ctifthenelse{\isnull{brand}}{Var('brand',\ROWID)}{brand} $$}
is translated into the SQL expression
\begin{lstlisting}
CASE WHEN brand IS NULL 
     THEN (SELECT value FROM best_guess_brand b 
           WHERE b.param1 = Product.ROWID) 
     ELSE brand END
\end{lstlisting}

\end{example}

To populate the best guess tables, Mimir simulates execution of the lens query, and identifies every variable instance that is used when constructing the lens' output.
The result of calling \texttt{getBestGuess} on the corresponding model is inserted into the best-guess table.
When a non-deterministic query is run, all references to $Var$ terms are replaced by nested lookup queries which read the values for $Var$ terms from  the corresponding best guess tables. 
As a further optimization, the in-lined lens query can also be pre-computed as a materialized view.

\subsubsection{Recovering Provenance}

This approach allows deterministic relational databases to directly evaluate
best-guess queries over C-Tables, eliminating the need for a shim query $\F$
to produce results.
However, the shim query also provides a form of provenance, linking 
individual results to the $Var$ terms that might affect them.  
Mimir's annotated cursors rely on this link to efficiently determine 
whether a result row or cell is uncertain and also when constructing explanation 
objects.

For inlining to be compatible with annotated cursors, three further changes 
are required:
(1) To retain the ability to quickly determine whether a given result
row or column is deterministic, result relations are extended with a 
`determinism' attribute for the row and for each column.
(2) To quickly construct explanation objects, we inject a provenance
marker into each result relation that can be used with the shim query $\F$ 
to quickly reconstruct any row or cell's full provenance.
(3) To count the number of potentially missing rows, we initiate a secondary
arity-estimation query that is evaluated off of the critical path.

\subsubsection{Result Determinism}
Recall from Example~\ref{ex:conditionalNonDeterminism} that expressions
involving conditionals, conjunctions, and disjunctions can create situations
where the determinism of a row or column is data dependent.  In the naive
execution strategy, these situations arise exclusively in the shim query $\F$
and can be easily detected.  As the first step towards recovering annotated 
cursors, we push this computation down into the query itself.  

Concretely, we rewrite a query $Q$ with schema $sch(Q) = \{a_i\}$ into a
new query $[[Q]]_{det}$ with schema $\{a_i, D_i, \phi\}$.  Each $D_i$
is a boolean-valued attribute that is true for rows where the corresponding
$a_i$ is deterministic.  $\phi$ is a boolean-valued attribute that is true for 
rows deterministically in the result set.  We refer to these two added sets
of columns as attribute- and row-determinism metadata, respectively. 
Query $[[Q]]_{det}$ is derived from the input query $Q$ by applying the operator 
specific rewrite rules described below, in a top-down fashion starting from the root operator of query $Q$.
%

\tinysection{Projection}
The projection rewrite relies on a variant of Algorithm~\ref{alg:isdet}, 
which rewrites columns according to the determinism of the input.
Consequently, the only change is that the column rewrite on line 5 replaces columns
with a reference to the column's attribute determinism metadata: \\[1mm]
\hspace*{5mm}4: \;\textbf{else if} E \textbf{is} $Column_i$ \textbf{then}\\
\hspace*{5mm}5: \;\;\;\;\;\;\textbf{return} $D_i$\\[1mm]
The rewritten projection is computed by extending the projection's output
with determinism metadata. Attribute determinism metadata is
computed using the expression returned by \texttt{isDet} and row determinism metadata is passed-through
unchanged from the input.
$$[[\pi_{a_i \leftarrow e_i}(Q)]]_{det} \mapsto \pi_{a_i \leftarrow e_i, D_i \leftarrow \texttt{isDet}(e_i), \phi \leftarrow \phi}([[Q]]_{det})$$

\tinysection{Selection}
Like projection, the selection rewrite makes use of \texttt{isDet}.  The selection
is extended with a projection operator that updates the row determinism 
metadata if necessary.
$$[[\sigma_{\psi}(Q)]]_{det} \mapsto \pi_{a_i \leftarrow a_i, D_i \leftarrow D_i, \phi \leftarrow \phi \wedge \texttt{isDet}(\psi)}(\sigma_{\psi}([[Q]]_{det}))$$

\tinysection{Cross Product}
Result rows in a cross product are deterministic if and only if both of their
input rows are deterministic.  Cross products are wrapped in a projection
operator that combines the row determinism metadata of both inputs,
while leaving the remaining attributes and attribute determinism metadata
intact.
$$[[Q_1\times Q_2]]_{det} \mapsto \pi_{a_i \leftarrow a_i, D_i \leftarrow D_i, \phi \leftarrow \phi_1 \wedge \phi_2}([[Q_1]]_{det} \times [[Q_2]]_{det})$$

\tinysection{Union}
Bag union already preserves the determinism metadata correctly and
does not need to be rewritten.
$$[[Q_1\cup Q_2]]_{det} \mapsto [[Q_1]]_{det} \cup [[Q_2]]_{det}$$

\tinysection{Relations}
The base case of the rewrite, once we arrive at a deterministic relation, we
annotate each attribute and row as being deterministic.
$$[[R]]_{det} \mapsto \pi_{a_i \leftarrow a_i, D_i \leftarrow \top, \phi \leftarrow \top}(R)$$

\tinysection{Optimizations}
These rewrites are quite conservative in materializing the full set of 
determinism metadata attributes at every stage of the query.  It is
not necessary to materialize every $D_i$ and $\phi$ if they can be
computed statically based solely on each operator's output.  For example, consider a given $D_i$ 
that is data-independent, as in a deterministic relation or an attribute 
defined by a $Var$ term.  $D_i$ has the same value for every row, and
can be factored out of the query.  A similar property holds for Joins
and Selections, allowing the projection enclosing the rewritten operator
to be avoided.

\subsubsection{Explanations}
Recall that explanation objects provide a way to analyze the non-determinism
in a given result row or cell.  Given a query $Q(D)$ and its normalized form
$\F(Q'(D))$, this analysis requires only $\F$ and the individual row in the 
output of $Q'(D)$ used to compute the row or cell being explained.

We now show how to construct a provenance marker during evaluation
of $Q$ and how to use this provenance marker to reconstruct the single
corresponding row of $Q'$.  The key insight driving this process is that
the normalization rewrites for cross product and union 
(Rewrites~\ref{rewrite:cross} and \ref{rewrite:union} in Figure~\ref{fig:normalFormReduction}) are isomorphic 
with respect to the data dependency structure of the query; $Q$ and 
$Q'$ both have unions and cross products in the same places.

As the basis for provenance markers, we use an implicit, unique per-row 
identifier attribute called \texttt{ROWID} supported by many popular 
database engines.  When joining two relations in the in-lined query, 
their \texttt{ROWID}s are concatenated (we denote string concatenation as $\circ$):
$$Q_1 \times Q_2 \mapsto \pi_{a_i \leftarrow a_i, \texttt{ROWID} \leftarrow \texttt{'('} \,\circ\, \texttt{ROWID}_1 \,\circ\, \texttt{')('} \,\circ\,\texttt{ROWID}_2 \,\circ\, \texttt{')'}}(Q_1 \times Q_2)$$
When computing a bag union, each source relation's \texttt{ROWID} is tagged
with a marker that indicates which side of the union it came from:
\begin{multline*}
Q_1 \cup Q_2 \mapsto \pi_{a_i \leftarrow a_i, \texttt{ROWID} \leftarrow \texttt{ROWID}\,\circ\,\texttt{'+1'}}(Q_1)\\
\cup \pi_{a_j \leftarrow a_j, \texttt{ROWID} \leftarrow \texttt{ROWID}\,\circ\, \texttt{'+2'}}(Q_2)
\end{multline*}
Selections are left unchanged, and projections are rewritten to pass 
the \texttt{ROWID} attribute through.

The method \texttt{unwrap}, summarized in Algorithm~\ref{alg:unwrap}, illustrates
how a symmetric descent through the deterministic component of
a normal form query and a provenance marker can be used to produce a 
single-row of $Q'$.  The descent unwraps the provenance marker, recovering
the single row from each join leaf used to compute the corresponding row
of $Q'$.  

\begin{algorithm}
\caption{$\texttt{unwrap}(Q', id)$}
\label{alg:unwrap}
\begin{algorithmic}
\REQUIRE{$Q'$: The deterministic component of a VG-RA normal form query.}
\REQUIRE{$id$: A \texttt{ROWID} from the inlined query $Q$ that was normalized into $\F(Q'(D))$.}
\ENSURE{A query to compute row $id$ of $Q'$}
\IF{$Q'$ \textbf{is} $\pi(Q_1)$}
   \RETURN $\pi(\texttt{unwrap}(Q_1, id))$
\ELSIF{$Q'$ \textbf{is} $\sigma(Q_1)$}
   \RETURN $\sigma(\texttt{unwrap}(Q_1, id))$
\ELSIF{$Q'$ \textbf{is} $Q_1\times Q_2$ \textbf{and} $id$ \textbf{is} \texttt{($id_1$)($id_2$)}}
   \RETURN $\texttt{unwrap}(Q_1, id_1) \times \texttt{unwrap}(Q_2, id_2)$
\ELSIF{$Q'$ \textbf{is} $Q_1\cup Q_2$ \textbf{and} $id$ \textbf{is} \texttt{$id_1$+1}}
   \RETURN $\texttt{unwrap}(Q_1, id_1)$
\ELSIF{$Q'$ \textbf{is} $Q_1\cup Q_2$ \textbf{and} $id$ \textbf{is} \texttt{$id_1$+2}}
   \RETURN $\texttt{unwrap}(Q_2, id_1)$
\ELSIF{$Q'$ \textbf{is} $R$}
   \RETURN $\sigma_{\texttt{ROWID} = id}(R)$
\ENDIF
\end{algorithmic}
\end{algorithm}


\subsection{Approach 3: Hybrid}
\label{sec:hybrid}

The first two approaches provide orthogonal benefits.  
The partitioning approach results in faster execution of queries over deterministic fragments of the data, as it is easier for the backend database query optimizer to take advantage of indexes already built over the raw data.
The inlining approach results in faster execution of queries over non-deterministic fragments of the data, as joins over non-deterministic values do not create a polynomial explosion of possible results.  
Our third and final approach is a simple combination of the two: Queries are first partitioned as in Approach 1, and then non-deterministic partitions are in-lined as in Approach 2.


\section{Experiments}
\label{sec:experiments}
We now summarize our the results of experimental analysis of the two optimizations 
presented in this paper.  We evaluate Virtual C-Tables under the classical 
normalization-based execution model, and partition-, inline-, and hybrid-optimized 
execution models.  All experiments are conducted using both SQLite as a backend, 
and a major commercial database termed DBX due to its licensing agreement.  
Mimir is implemented in Scala and Java.  Measurements presented are for Mimir's 
textual front-end.
All experiments were run under RedHat Enterprise Linux 6.5 on a 16 core 2.6 GHz 
Intel Xeon server with 32 GB of RAM and a 4-disk 900 GB RAID5 array.  
Mimir and all database backends were hosted on the same machine to avoid including 
network latencies in measurements.  
Our experiments demonstrate that: (1) Virtual C-Tables scale well, (2) Virtual C-
Tables impose minimal overhead compared to deterministic evaluation, and (3) 
Hybrid evaluation is typically optimal.

\subsection{Experimental Setup}
Datasets were constructed using TPC-H~\cite{tpch}'s dbgen with scaling factors 1 (1 GB) and 0.1 (100 MB).  
To simulate incomplete data that could affect join predicates, we randomly replaced a percentage of foreign key references in the dataset with \texttt{NULL} values.  We created domain constraint repair lenses over the damaged relations to ``repair'' these \texttt{NULL} values as non-materialized views.
As a query workload, we used TPC-H Queries 1, 3, 5, and 9 modified in two ways.  First, all relations used by the query were replaced by references to the corresponding domain constraint repair lens.  Second, Mimir does not yet include support for aggregation.  Instead we measured the cost of enumerating the set of results to be aggregated by stripping out all aggregate functions and computing their parameter instead\footnote{The altered queries can be found at {\small \url{https://github.com/UBOdin/mimir/tree/master/test/tpch_queries/noagg}}}.  Execution times were capped at 30 minutes.

We experimented with two different backend databases: \textbf{SQLite} and a major commercial database \textbf{DBX}.  We tried four different evaluation strategies: \textbf{Classic} is the naive, normalization-based evaluation strategy, while \textbf{Partition}, \textbf{Inline}, and \textbf{Hybrid} denote the optimized approaches presented in Sections \ref{sec:partition}, \ref{sec:inline}, and \ref{sec:hybrid} respectively.  \textbf{Deterministic} denotes the four test queries run directly on the backend databases with un-damaged data, and serves as an \textit{lower bound} for how fast each query can be run.

\subsection{Comparison}

\begin{figure*}
  \begin{subfigure}[t]{0.5\textwidth}
    \centering
    \includegraphics[width=0.9\textwidth]{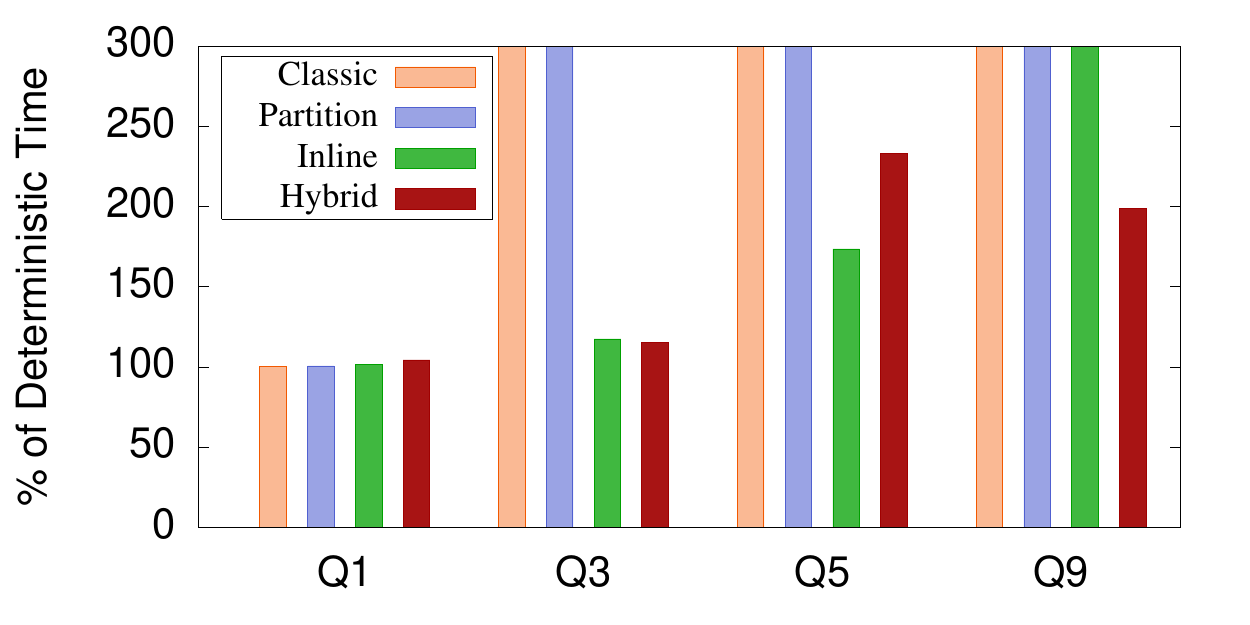}\\[-3mm]
    \caption{TPC-H SF 0.1 (100 MB)}
    \label{fig:sqlite:100m}
  \end{subfigure}
  \begin{subfigure}[t]{0.5\textwidth}
    \centering
    \includegraphics[width=0.9\textwidth]{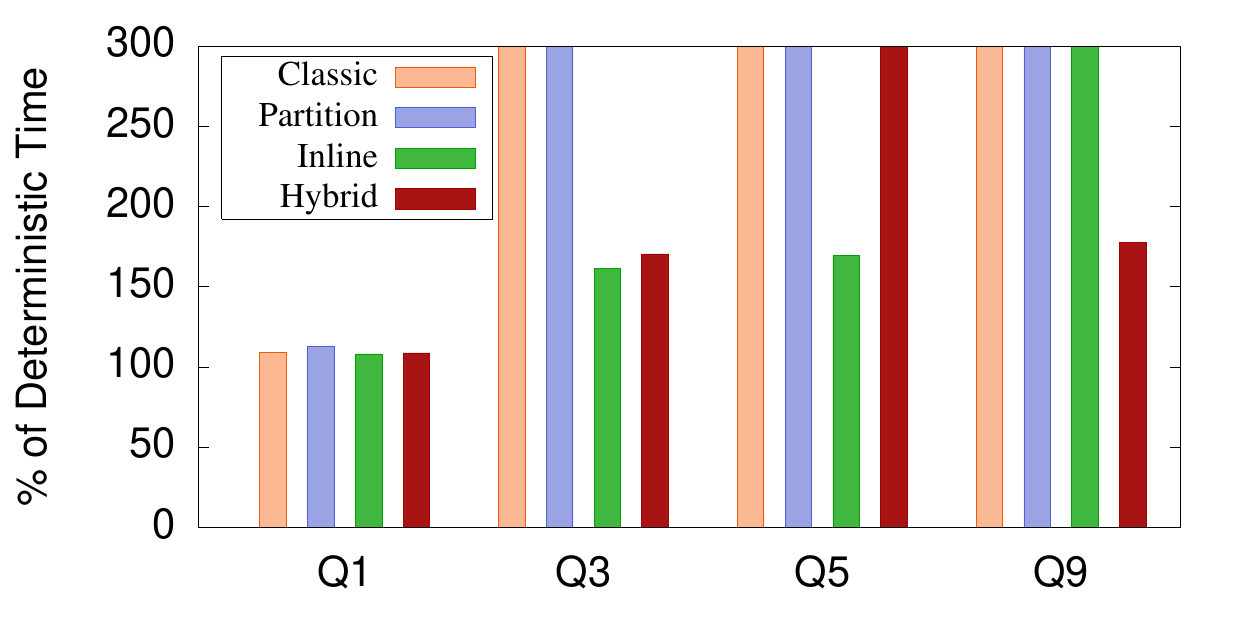}\\[-3mm]
    \caption{TPC-H SF 1 (1 GB)}
    \label{fig:sqlite:1g}
  \end{subfigure}
  \caption{Performance of Mimir running over SQLite as a percent of deterministic query execution time.}
  \label{fig:sqlite}
\end{figure*}

\begin{figure*}
  \begin{subfigure}[t]{0.5\textwidth}
    \centering
    \includegraphics[width=0.9\textwidth]{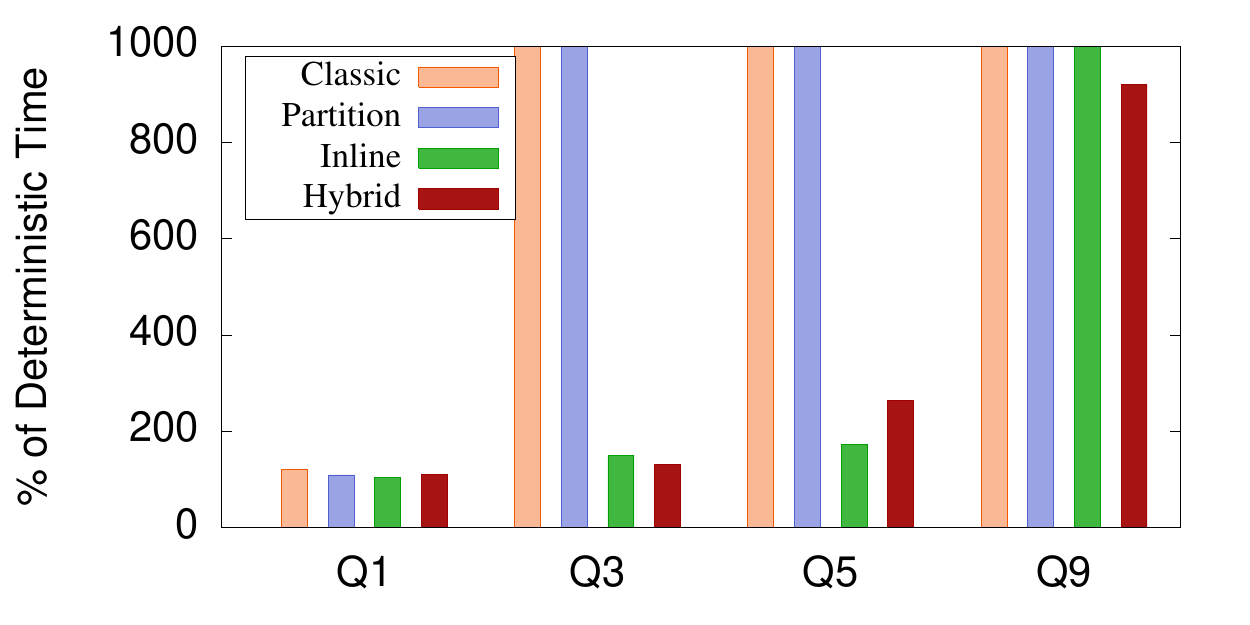}\\[-3mm]
    \caption{TPC-H SF 0.1 (100 MB)}
    \label{fig:dbx:100m}
  \end{subfigure}
  \begin{subfigure}[t]{0.5\textwidth}
    \centering
    \includegraphics[width=0.9\textwidth]{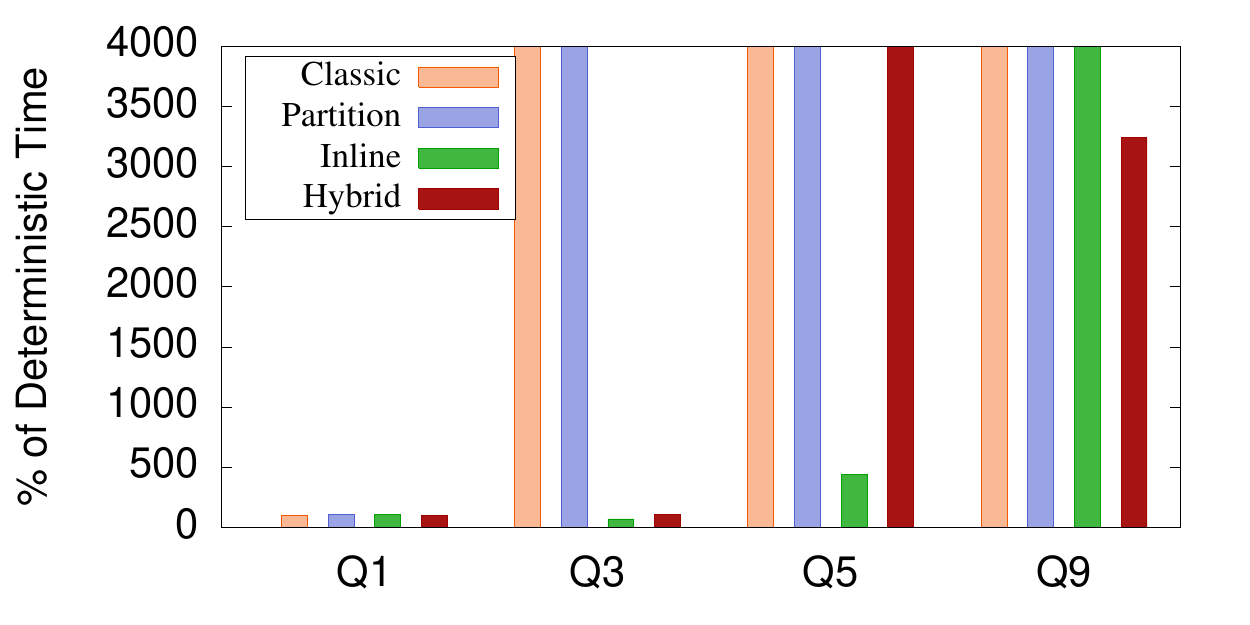}\\[-3mm]
    \caption{TPC-H SF 1 (1 GB)}
    \label{fig:dbx:1g}
  \end{subfigure}
  \caption{Performance of Mimir running over DBX as a percent of deterministic query execution time.}
  \label{fig:dbx}
\end{figure*}

Figures~\ref{fig:sqlite} and \ref{fig:dbx} show the performance of Mimir running over SQLite and DBX, respectively.  The graphs show Mimir's overhead relative to the equivalent deterministic query.  

\tinysection{Table scans are unaffected by Mimir}  Query 1 is a single-table scan.  In all configurations, Mimir's overhead is virtually nonexistent.

\tinysection{Partitioning accelerates deterministic results}  Query 3 is a 3-way foreign-key lookup join.  Under naive partitioning, completely deterministic partitions are evaluated almost immediately.  Even with partitioning, non-deterministic subqueries still need to be partly evaluated as cross products, and partitioning times out on all remaining queries.

\tinysection{Partitioning can be harmful}  Query 5 is a 6-way foreign-key lookup join where Inline performs better than Hybrid.  Each foreign-key is dereferenced in exactly one condition in the query, allowing Inline to create a query with a plan that can be efficiently evaluated using Hash-joins.  The additional partitions created by Hybrid create a more complex query that is more expensive to evaluate.

\tinysection{Partitioning can be helpful}  Query 9 is a 6-way join with a cycle in its join graph.  Both \texttt{PARTSUPP} and \texttt{LINEITEM} have foreign key references that must be joined together.  Consequently, Inlining creates messy join conditions that neither backend database evaluates efficiently.  Partitioning results in substantially simpler nested queries that both databases accept far more gracefully.


\section{Related Work}

The design of Mimir draws on a rich body of literature, spanning the  areas probabilistic databases, model databases, provenance, and data cleaning.  We now relate key contributions in these areas on which we have based our efforts.

\tinysection{Incomplete Data} 
Enabling queries over incomplete and probabilistic data has been an area of active research for quite some time.  Early research in the area includes \texttt{NULL} values~\cite{DBLP:journals/tods/Codd79}, the C-Tables data model~\cite{DBLP:journals/jacm/ImielinskiL84} for representing incomplete information, and research on fuzzy databases~\cite{ZEMANKOVA1985107}.  The C-Tables representation used by Mimir has been linked to both probability distributions and provenance through so-called PC-Tables~\cite{DBLP:journals/debu/GreenT06} and Provenance Semirings~\cite{Green:2012}, respectively.  
These early concepts were implemented through a plethora of probabilistic database systems.  
Most notably, MayBMS~\cite{DBLP:conf/sigmod/HuangAKO09} employs a simplification of C-Tables called U-Relations that does not rely on labeled nulls, and can be directly mapped to a deterministic relational database.  However, U-Relations can only encode uncertainty described by finite discrete distributions (e.g., Bernoulli), while VC-Tables can support continuous, infinite distributions (e.g., Gaussian).
Other probabilistic database systems include MCDB~\cite{jampani2008mcdb}, Sprout~\cite{DBLP:conf/sigmod/FinkHOR11}, Trio~\cite{DBLP:conf/vldb/AgrawalBSHNSW06}, Orion 2.0~\cite{Singh:2008:ONS:1376616.1376744}, Mystiq~\cite{DBLP:conf/sigmod/BoulosDMMRS05}, Pip~\cite{kennedy2010pip}, Jigsaw~\cite{DBLP:conf/sigmod/KennedyN11}, and numerous others.  
These systems all require heavyweight changes to the underlying database engine, or an entirely new database kernel.  By contrast, Mimir is an external component that attaches to an existing deployed database, and can be trivially integrated into existing deterministic queries and workflows.

\tinysection{Model Databases}
A specialized form of probabilistic databases focus on representing structured models such as graphical models or markov processes as relational data.  These types of databases exploit the structure of their models to accelerate query evaluation.  Systems in this space include BayesStore~\cite{Wang:2008:BML:1453856.1453896}, MauveDB~\cite{DBLP:conf/sigmod/DeshpandeM06}
Lahar~\cite{4812407}, and SimSQL~\cite{Cai:2013:SDM:2463676.2465283}.  In addition to defining semantics for querying models, work in this space typically explores techniques for training models on deterministic data already in the database.  The  Mimir system treats lens models as black boxes, ignoring model structure.  It is likely possible to incorporate model database techniques into Mimir.  We leave such considerations as future work.

\tinysection{Provenance}
Provenance (sometimes referred to as lineage) describes how the outputs of a computation are derived from relevant inputs.  Provenance tools provide users with a way of quickly visualizing and understanding how a result was obtained, most commonly as a way to validate outliers, better understand the results, or to diagnose errors.  Examples of provenance systems include a general provenance-aware database called Trio~\cite{DBLP:conf/vldb/AgrawalBSHNSW06}, a collaborative data exchange system called Orchestra~\cite{DBLP:journals/debu/GreenKIT10}, and a generic database provenance middleware called GProM which also supports updates and transactions~\cite{arab2014generic,arab2014reenacting}.
It has been shown that certain types of provenance can encode C-Tables~\cite{Green:2012}.  It is this connection that allows Mimir to provide reliable feedback about sources of uncertainty in the results.  The VG-Relational algebra used in Mimir creates symbolic expressions that act as a form of provenance similar to semiring provenance~\cite{Green:2012} and its extensions to value expressions (aggregation~\cite{Green:2012}) and updates~\cite{arab2014reenacting}.

\tinysection{Data Curation}
In principle, it is useful to query uncertain or incomplete data directly.  However, due to the relative complexity of declaring uncertainty upfront, it is still typical for analysts to validate, standardize, and merge data \textit{before} importing it into a data management system for analysis.  Common problems in the space of data curation include entity de-duplication~\cite{Elmagarmid:2007:DRD:1191547.1191739,Panse:2013:IHU:2435221.2435225,Sarawagi:2002:IDU:775047.775087}, interpolation~\cite{DBLP:conf/sigmod/DeshpandeM06,Mayfield:2010:EDA:1807167.1807178}, schema matching~\cite{bernstein2011generic,DBLP:conf/icde/McCannSD08,lee2007etuner,rahm2001survey}, and data fusion.  
Mimir's lenses each implement a standard off-the-shelf curation heuristic.  These heuristics usually require manual tuning, validation, and refinement.  By contrast, in Mimir these difficult, error-prone steps can be deferred until query-time, allowing analysts to focus on the specific cleaning tasks that are directly relevant to the query results at hand.  

Other systems for simplifying or deferring curation exist.  For example, DataWrangler~\cite{Kandel:2011:WIV:1978942.1979444} creates a data cleaning environment that uses visualization and predictive inference to streamline the data curation process.  Similar techniques could be used in Mimir for lens creation, streamlining data curation even further.  Mimir can also trace its roots to on-demand cleaning systems like Paygo~\cite{DBLP:conf/sigmod/JefferyFH08}, CrowdER~\cite{Wang:2012:CCE:2350229.2350263}, and GDR~\cite{Yakout:2011:GDR:1952376.1952378}.  In contrast to Mimir, these systems each focus on a specific type of data cleaning: Duplication and Schema Matching, Deduplication, and Conditional Functional Dependency Repair, respectively.  Mimir provides a general curation framework that can incorporate the specialized techniques used in each of these systems.

\tinysection{Uncertainty Visualization}
Visualization of uncertain or incomplete data arises in several domains.  As already noted, DataWrangler~\cite{Kandel:2011:WIV:1978942.1979444} uses visualization to help guide users to data errors.  MCDB~\cite{jampani2008mcdb} uses histograms as a summary of uncertainty in query results.  Uncertainty visualization has also been studied in the context of Information Fusion~\cite{Todoran:2015:MEI:2788681.2744205,bisantz1999human}.  Mimir's explanation objects are primitive by comparison, and could be extended with any of these techniques.


\section{Conclusions}
We presented Mimir, an on-demand data curation system based on a novel type of probabilistic data curation operators called Lenses. The system sits as a shim layer on-top of a relational DBMS backend - currently we support SQLite and a commercial system. Lenses encode the result of a curation operation such as domain repair or schema matching as a probabilistic relation. The driving force behind Mimir's implementation of Lenses are VC-Tables which are a representation of uncertain data that cleanly separates the existence of uncertainty from a probabilistic model for the uncertainty. 
This enables efficient implementation of queries over lenses by outsourcing the deterministic component of a query to the DBMS. Furthermore, the symbolic expressions used by VC-Tables to represent uncertain values and conditions act as a type of provenance that can be used to explain how uncertainty effects a query result.
In this paper we have introduced several optimizations of this approach that 1) push part of the probabilistic computation into the database (we call this \textit{inlining}) without loosing the ability to generate explanations and 2) partitioning a query based on splitting selection conditions such that some fragments can be evaluated deterministically or can benefit from available indexes.
In future work we will investigate cost-based optimization techniques for lens queries and using the probabilistic model for uncertainty in the database to exclude rows that are deterministically not in the result (e.g., if a missing brand value is guaranteed to be either Apple or Samsung, then this value does not fulfill a condition $brand = Sony$).


{\small
\bibliographystyle{plain}
\bibliography{main}
}

\end{document}